\documentclass[conference]{IEEEtran}

\IEEEoverridecommandlockouts                              

\usepackage[utf8]{inputenc}
\usepackage[T1]{fontenc}
\usepackage{graphicx} 
\graphicspath{ {./images/} }

\usepackage{listings}
\usepackage{color}
\usepackage{cite}
\usepackage{float}

\definecolor{dkgreen}{rgb}{0,0.6,0}
\definecolor{gray}{rgb}{0.5,0.5,0.5}
\definecolor{mauve}{rgb}{0.58,0,0.82}

\usepackage[rightcaption]{sidecap}

\usepackage{wrapfig}
\usepackage{graphics} 
\usepackage{epsfig} 
\usepackage{mathptmx} 
\usepackage{mathptmx} 
\usepackage{amsmath} 
\usepackage{amssymb}  

\title{\LARGE \bf
Generating Light-based Fingerprints for Indoor Localization
}

\author{
    \IEEEauthorblockN{$^{1*}$Hsun-Yu Lee, $^{1*}$Jie Lin, $^2$Fang-Jing Wu}
    \IEEEauthorblockA{$^1$Department of Information Management, National Taiwan University}
    \IEEEauthorblockA{$^*$Corresponding author: b11705022@ntu.edu.tw, b11705048@ntu.edu.tw } 
    \thanks{*Authors contributed equally to this work.}
  \thanks{This work was presented at the 2024 MC \& WASN Conference.}%
  \thanks{This paper was a candidate for the Best Paper Award.}
    \IEEEauthorblockA{fangjing@csie.ntu.edu.tw}
}

\begin{document}

\maketitle
\thispagestyle{empty}
\pagestyle{empty}

\begin{abstract}
Accurate indoor localization underpins applications ranging from wayfinding and emergency response to asset tracking and smart–building services. Radio‑frequency solutions (e.g., Wi‑Fi, RFID, UWB) are widely adopted but remain vulnerable to multipath fading, interference, and uncontrollable coverage variation. We explore an orthogonal modality—visible light communication (VLC)—and demonstrate that the spectral signatures captured by a low‑cost AS7341 sensor can serve as robust location fingerprints.

We introduce a two‑stage framework that (i) trains a multi–layer perceptron (MLP) on real spectral measurements and (ii) enlarges the training corpus with synthetic samples produced by \emph{TabGAN}. The augmented dataset reduces the mean localization error from \mbox{62.9~cm} to \mbox{49.3~cm}—a \textbf{20\%} improvement—while requiring only 5{\%} additional data‑collection effort. Experimental results obtained on 42 reference points in a U‑shaped laboratory confirm that GAN‑based augmentation mitigates data‑scarcity issues and enhances generalization.

\end{abstract}

\section{\textbf{Introduction}}
Indoor positioning systems (IPS) enable seamless navigation, emergency dispatch, asset management, and emerging location–aware services. Conventional IPS solutions exploit radio signals such as Wi‑Fi RSSI, Bluetooth beacons, and UWB pulses, but their accuracy deteriorates in richly scattering environments and their coverage is hard to bound. VLC‑based approaches, in contrast, benefit from line‑of‑sight propagation and the ubiquitous presence of LED lighting.

The work in \cite{7408994} focuses on an indoor localization system using visible light communication, while many technologies implement indoor localization with different signal sources, such as Wi-Fi, Radio Frequency Identification (RFID), Wireless Sensor Networks (WSN), Bluetooth, and Ultra Wideband (UWB) technology. The above methods use radio waves to transmit signals. However, radio signals are not reliable because of signal fluctuation. Also, it is difficult to control the radio coverage due to signal variations. Therefore, the above radio-based methods lead to uncertain data availability. 

Due to the aforementioned vulnerability of radio signals, our work uses visible light communication to solve the problem. We collect the light spectrum patterns using light sensors. The spectral data of visible light is used to train a DNN model for estimating the current position of sensing device. 

However, the process of data collection can be time-consuming and costly. We augment the data with the help of GANs to generate additional fake data. Recently, DNN-based RSSI (Received Signal Strength Indicator) fingerprints have been paid attention to enhance localization performance. However, DNN-based approaches require a large amount of training data. To solve the problem, \cite{Njima2021-SGAI-loc} suggests using Generative Adversarial Networks (GANs) for RSSI data augmentation to generate fake RSSI data based on a small set of real data for training. 

Generally, the deep learning process consists of two models: a discriminator and a generator. We use Multi-Layer Perceptron (MLP) as the discriminative model for location classification and GANs as the generative model for generating synthetic data. Our approach consists of with the steps below:

\begin{enumerate}
    \item Collecting spectral data from our laboratory, and training a localization MLP model based on the real data.
    \item Augmenting the spectral data using TabGAN. \cite{ashrapov2020-tabular}
    \item Giving the fake data pseudo-labels by the MLP model in step 1.
    \item Training a new model with mixed data (real data and fake data).
    \item Using the MLP model we trained in step 4 to evaluate the result.
\end{enumerate}

Further details will be provided in the forthcoming sections of the paper.

\begin{figure}
    \centering
    \includegraphics[width=1.0\linewidth, height=0.25\textheight]{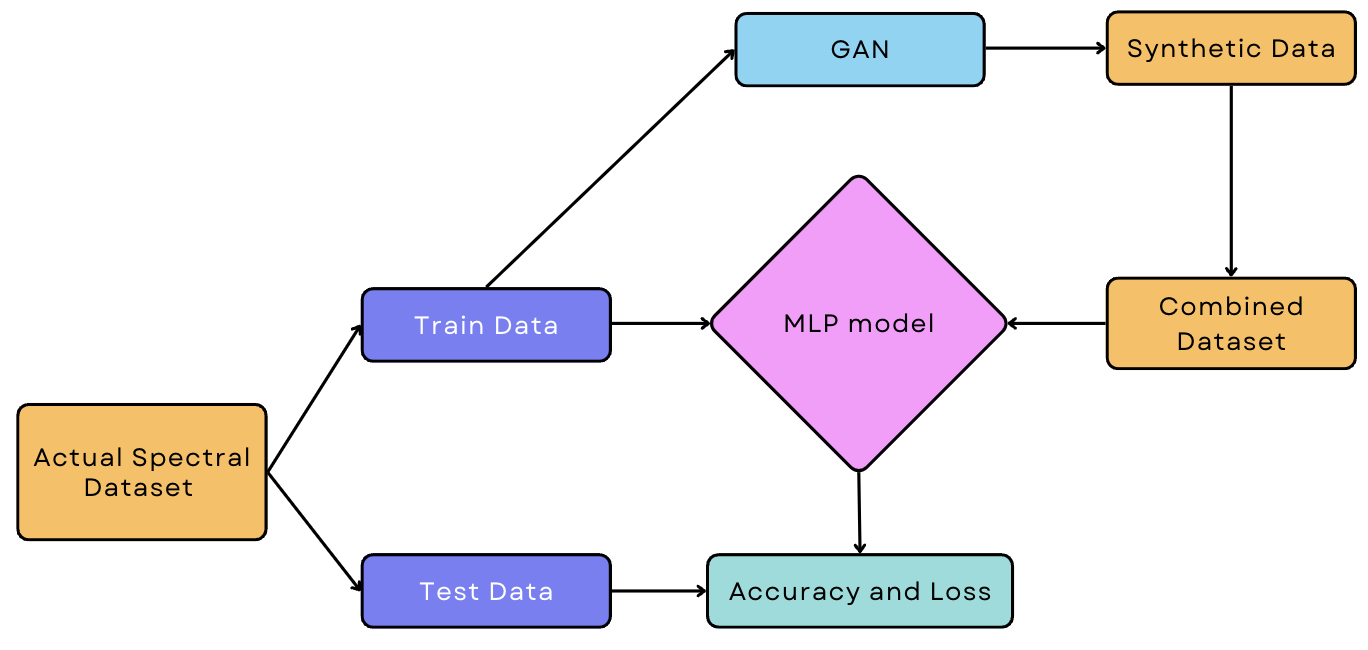}
    \caption{Model architecture.}
    \label{fig:enter-label}
\end{figure}

\section{\textbf{Data Collection}}

\subsection{The Sensor}

\begin{figure}[h]
    \centering
    \includegraphics[width=0.25\textwidth]{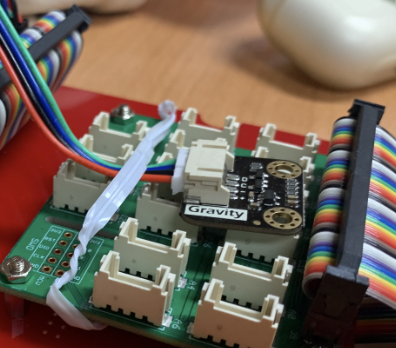}
    \caption{AS7341 Sensor.}
    \label{fig:mesh1}
\end{figure}

We employ the AMS AS7341 spectral sensor\,\cite{ashrapov2020-tabular} integrated with a Raspberry~Pi. The device outputs eight visible‑light channels (F1–F8), one near‑infrared channel, one flicker channel, and one clear channel at 1~Hz. Figure~\ref{fig:mesh1} depicts the sensor module.

\begin{figure}[ht]
    \centering
    \includegraphics[width=0.5\textwidth, height=0.3\textheight]{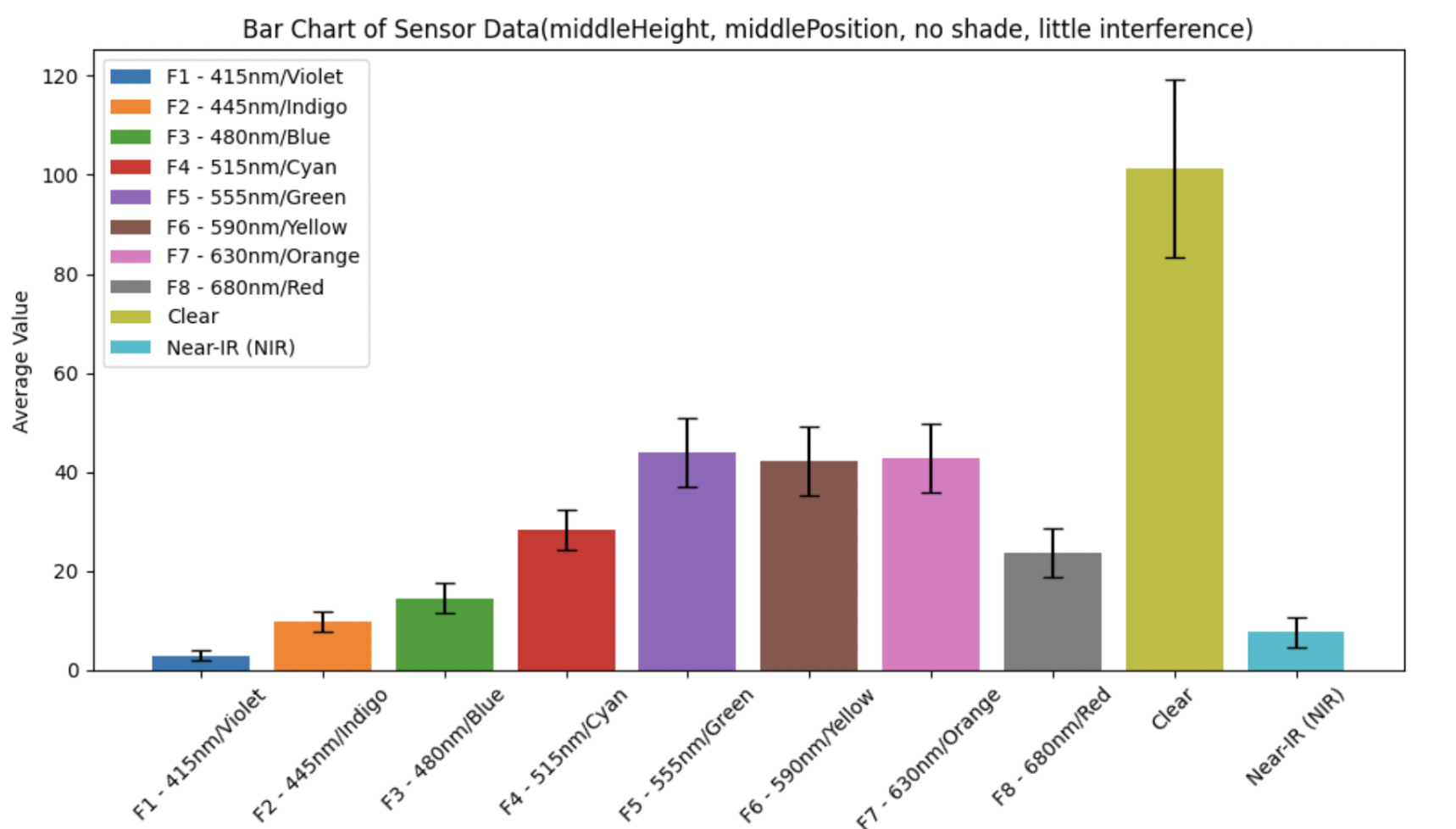}
    \caption{Spectral data: Jie's desk.
    x-axis: channels. y-axis: light intensity (unit).} 
    \label{fig:mesh2}
\end{figure}

\begin{figure}[ht]
    \centering
    \includegraphics[width=0.5\textwidth, height=0.3\textheight]{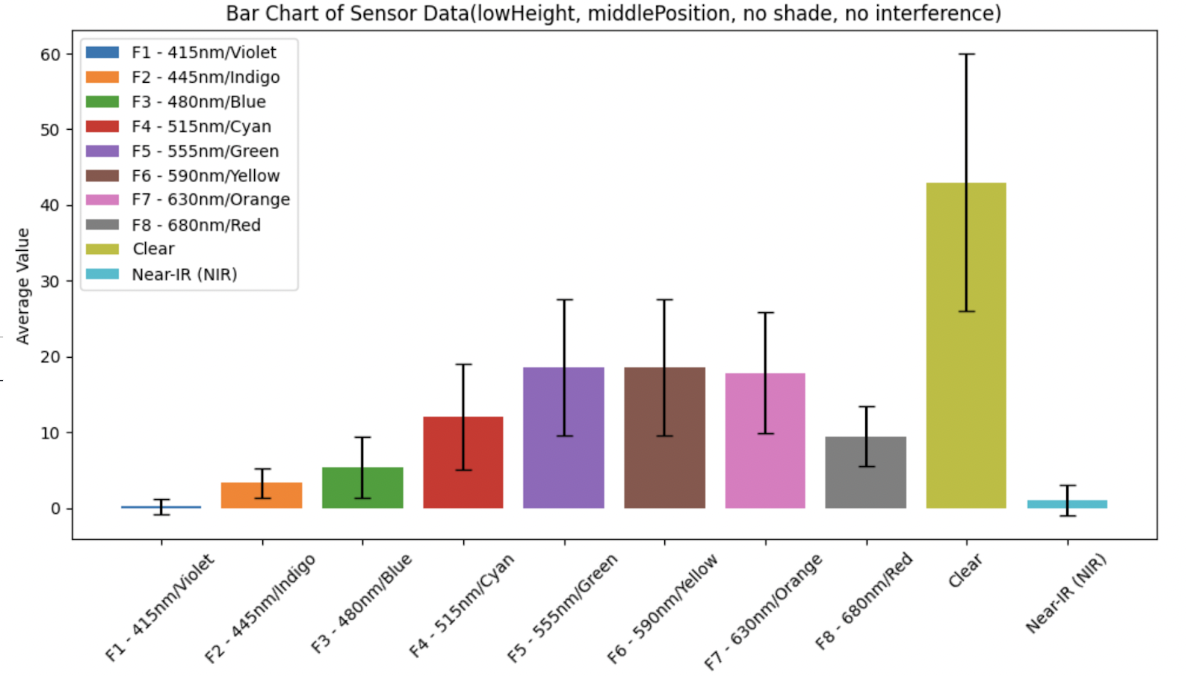}
    \caption{Spectral data: On the ground (below Jie's desk). x-axis: channels. y-axis: light intensity (unit).}
    \label{fig:mesh3}
\end{figure}

\begin{figure}[ht]
    \centering
    \includegraphics[width=0.5\textwidth, height=0.3\textheight]{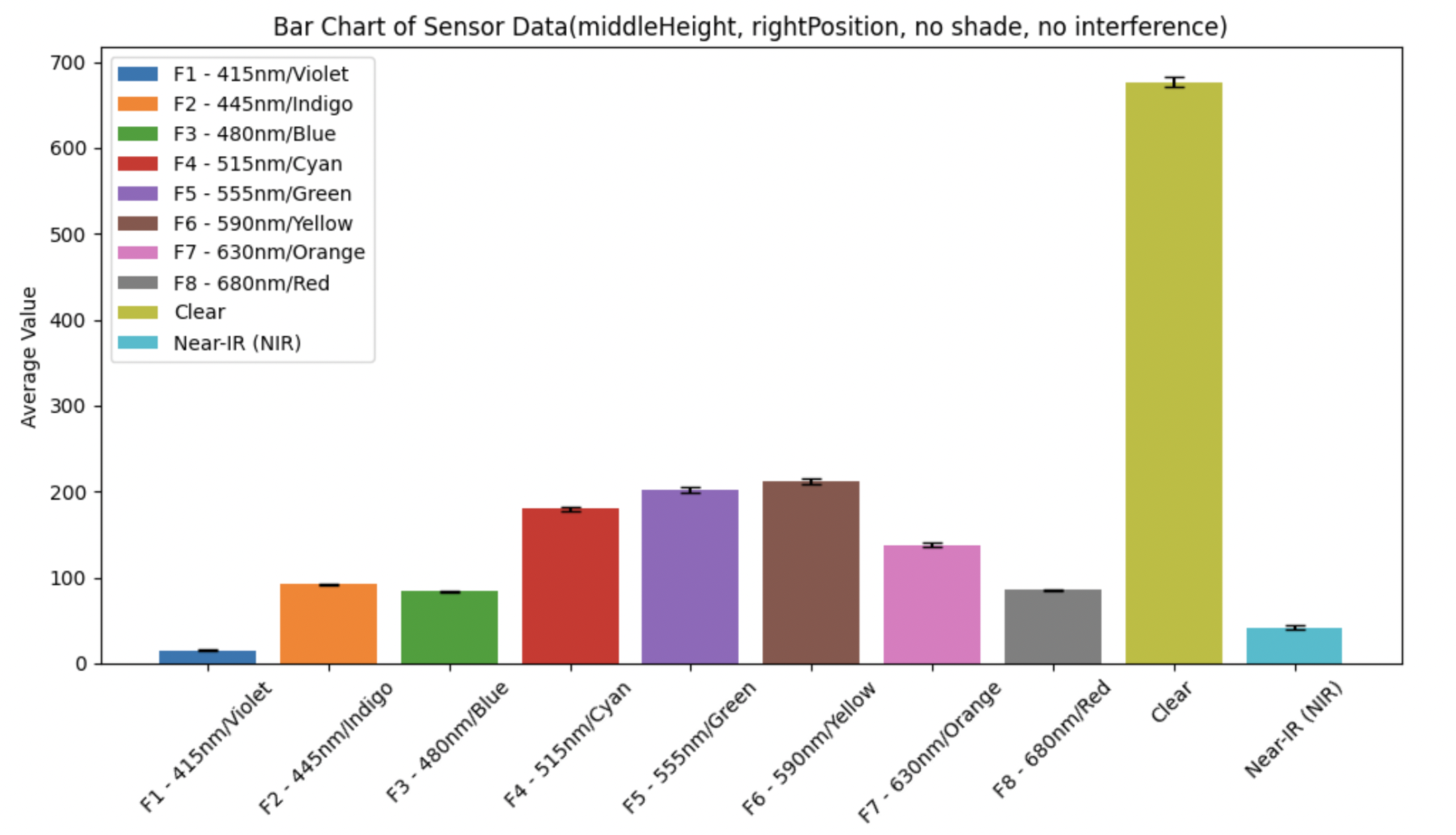}
    \caption{Spectral data: To the right of Jie's desk. x-axis: channels. y-axis: light intensity (unit).}
    \label{fig:mesh4}
\end{figure}

After we get the sensor, we would like to test whether it is suitable for us to use the spectral pattern as the fingerprint for indoor localization. We placed the sensor at different locations in the testing environment. For each location, we set the frequency to record one result per second, and then average the results after collecting 100 readings.
We placed the sensors at three different locations: in the middle of the desk (Fig.~\ref{fig:mesh2}), right below the desk (Fig.~\ref{fig:mesh3}), and 1 meter to the right of the middle of the desk (Fig.~\ref{fig:mesh4}). The distance between the three locations is relatively small since we would like to test whether the sensor can receive the change of patterns even though the sensor was just slightly moved.


According to Fig.~\ref{fig:mesh2}, Fig.~\ref{fig:mesh3}, Fig.~\ref{fig:mesh4}, not only does the spectral information from the three locations show different patterns, but the intensity also varies. Therefore, we believe that AS7341 is able to capture distinct spectral signatures associated with different locations. This evaluation will inform us whether the sensor's spectral pattern can serve as a reliable fingerprint for indoor localization purposes.

\subsection{\textbf{Experiment Environment}}

\begin{figure}[h]
    \centering
    \includegraphics[width=0.4\textwidth]{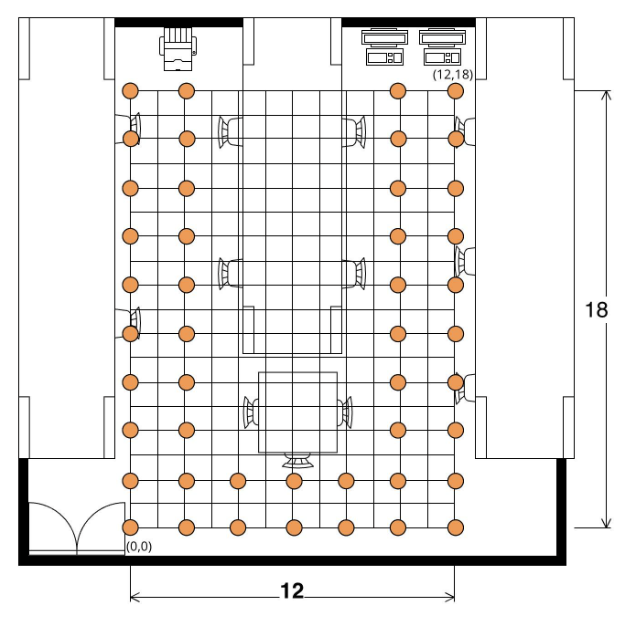}
    \caption{Data collection points in the laboratory.}
    \label{fig:mesh5}
\end{figure}

Measurements were recorded in a U‑shaped laboratory (\mbox{6~m}~\(\times\)~\mbox{8~m}) under static lighting. Forty‑two reference points (RP) were marked on the floor (Figure~\ref{fig:mesh5}). At each RP, the sensor remained stationary for 30~s and sampled at 4~Hz, producing 5040 spectra in total. Background objects and ambient luminosity were kept constant.

When we collected the data, we ensured the following requirements were met:
\begin{itemize}
    \item Background: No changes in the arrangement of furniture, tables, and appliances. No people around.
    \item When collecting test data, the sensor stayed at each marked point for 30 seconds. The sensor collects data at a frequency of four times per second, resulting in a total of problems rows of data.
\end{itemize}
In conclusion, we collected a total of 5040 rows of spectral data from 42 points in the room.

\begin{figure}[h]
    \centering
    \includegraphics[width=0.4\textwidth]{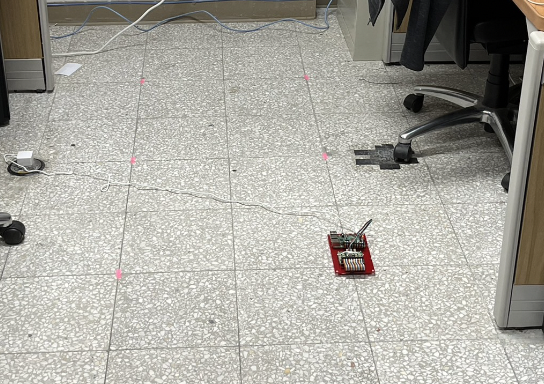}
    \caption{The sensor was receiving data in the laboratory.}
    \label{fig:mesh6}
\end{figure}

\section{\textbf{First Supervised MLP Model}}    

After retrieving the data, we then construct our first Multi-Layer Perceptron (MLP) model to predict the location using data from light signals. There are several methods we've tried, such as CNN, but it doesn't turn out as good is MLP.

\subsection{Model Architecture}

The model architecture is based on a fully connected neural network implemented using PyTorch. The network consists of multiple layers with ReLU activation functions and dropout for regularization. The architecture is optimized using Optuna for hyperparameter tuning.

\subsection{Performance}
The validation results are illustrated in ~\ref{fig:enter-label}. Each blue dot represents the ground-truth value y, while its corresponding red dot denotes the model’s prediction for the same x. On the right side of the room, the model performs well: red dots cluster tightly around the blue ones. In contrast, on the left side, the MLP fails to capture the underlying pattern, and the red dots are widely scattered. The final test loss is 62.93cm.

\begin{figure}[H]
    \centering
    \includegraphics[width=0.9\linewidth, height=0.33\textheight]{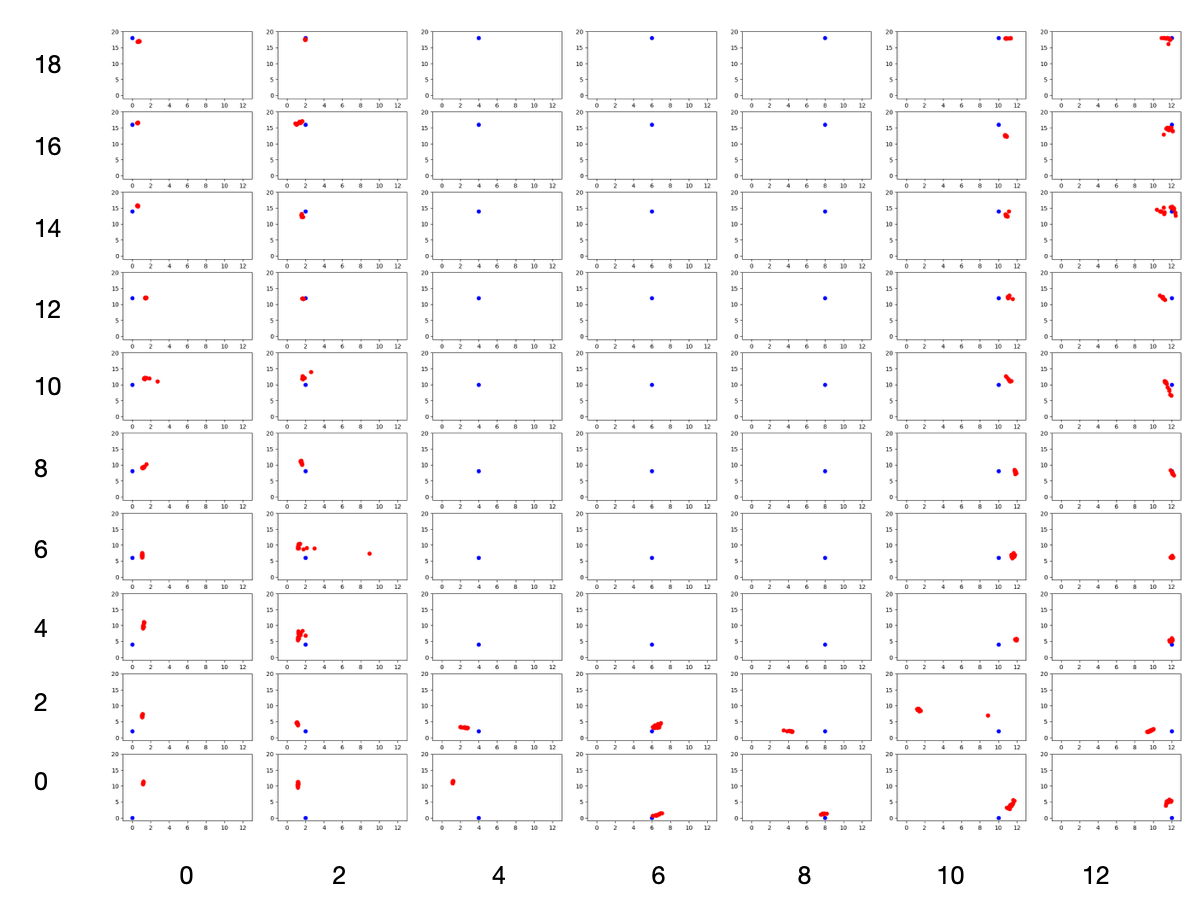}
    \caption{Performance of the model.}
    \label{fig:enter-label}
\end{figure}

So, in order to improve the performance, we then try to produce more diverse data with GAN to expand our dataset.

\section{\textbf{Generating Synthetic Spectral Data with GANs}}

Collecting spectral fingerprints is both costly and time-consuming. To enlarge our dataset without additional measurements, we adopt a Tabular GAN (TGAN) \cite{ashrapov2020-tabular} to generate synthetic spectra that mimic the distribution of the real data, thereby improving model robustness and accuracy.

\subsection{Pre-processing}

The measured spectra were reformatted into two files:  
(i) a full dataset containing each sample’s \(x\)-\(y\) coordinates, and  
(ii) a coordinate-free dataset containing only the spectral vectors.
We train TGAN on the coordinate-free dataset so that the generator focuses purely on spectral characteristics.

\subsection{GAN Training and Sampling}

Tabular GAN learns the joint distribution of all numerical variables and can therefore sample realistic frequency vectors. After training on the real spectra collected in our lab, the model was used to generate 6,000 synthetic samples.

\subsection{Distribution Analysis}

\begin{figure}
    \centering
    \includegraphics[width=\linewidth]{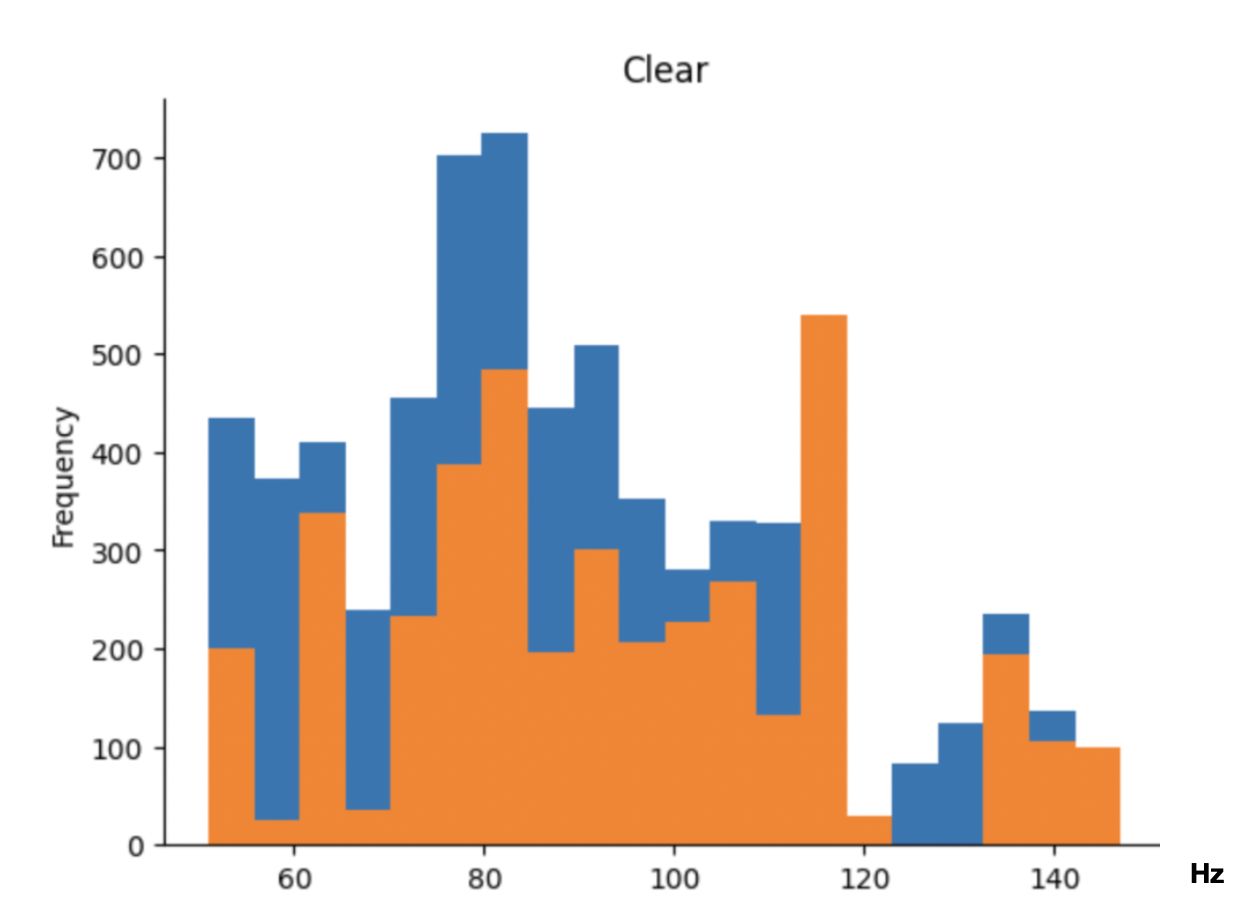}
    \caption{Intensity distribution on the “Clear” channel. Blue: real data; orange: synthetic data.}
    \label{fig:clear-dist}
\end{figure}

\begin{figure}
    \centering
    \includegraphics[width=\linewidth]{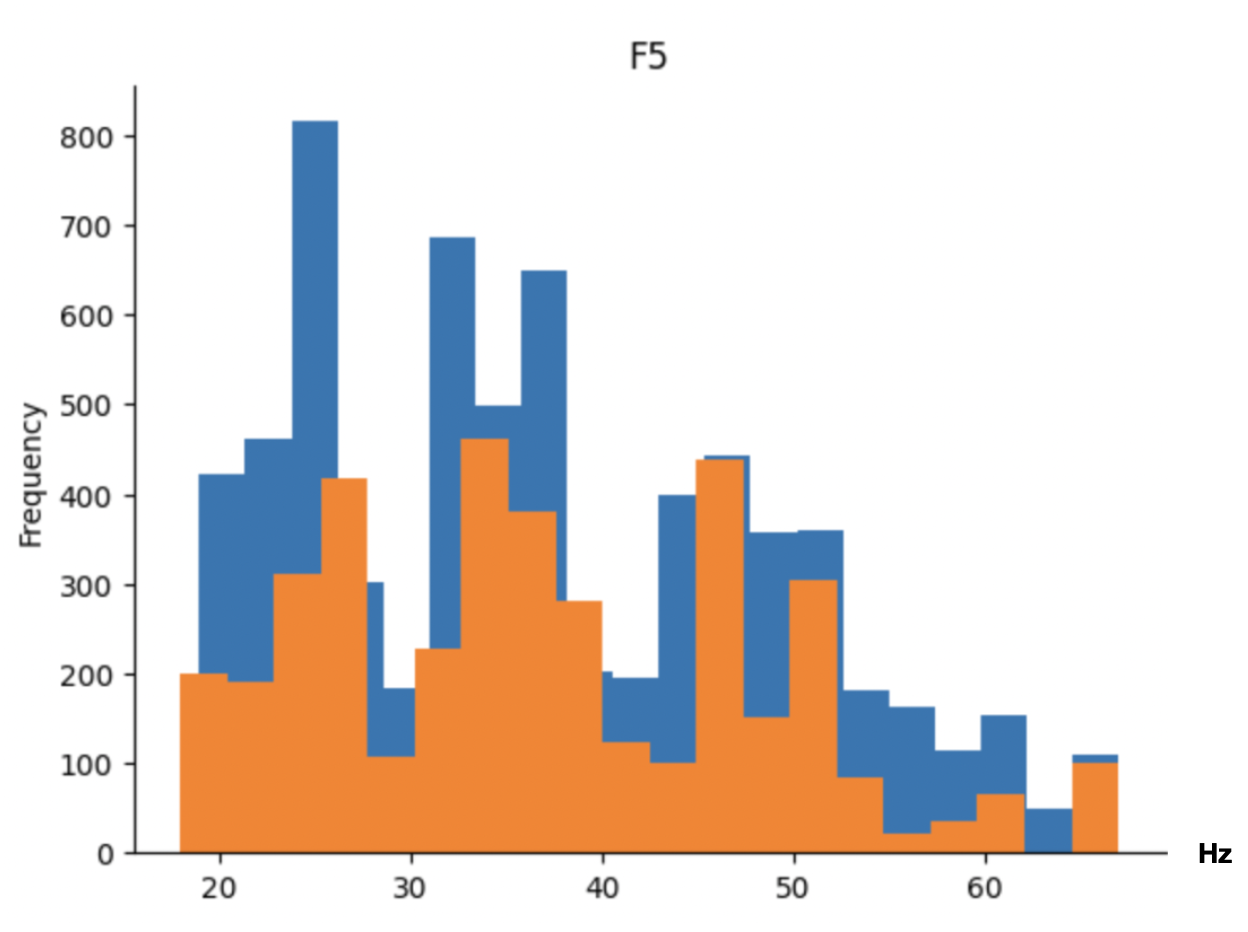}
    \caption{Intensity distribution on the “F5” channel. Blue: real data; orange: synthetic data.}
    \label{fig:f5-dist}
\end{figure}

To verify fidelity, we compare intensity histograms of real and synthetic data on two representative channels (“Clear” and “F5”). For each channel, we bin the intensities and count occurrences; the resulting histograms are shown in Figs. ~\ref{fig:clear-dist}, ~\ref{fig:f5-dist}

The close alignment of the two histograms indicates that TGAN successfully captured the underlying spectral distribution and produced high-fidelity synthetic data suitable for augmenting our training set.

\section{\textbf{Training With the Pseudo Data}}

After generating data using a Generative Adversarial Network (GAN), we integrate this synthetic data into our original model to predict the location values. By analyzing the results shown in Fig 8, we noticed that the performance of the location predictions varied. Specifically, some locations did not perform as expected, prompting us to increase the number of data points generated at those locations to improve the model's accuracy.

We generated a total of 6,000 data points using the GAN. However, 1,002 of these points were discarded because their pseudo labels were out of bounds. The remaining data predominantly clustered around the left aisle and the door. This targeted data augmentation was intended to address areas where the model's predictions were weaker.

Subsequently, we retrained the model using the same architecture as described in Section \textbf{First Supervised MLP Model}. We also performed additional hyperparameter tuning to optimize the model's performance with the augmented dataset.

\begin{figure}
    \centering
    \includegraphics[width=1.0\linewidth]{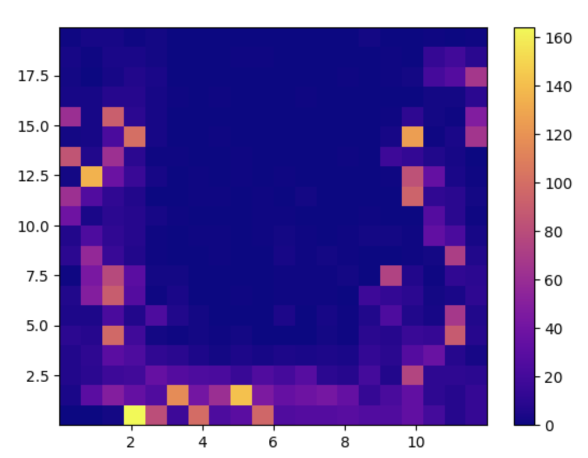}
    \caption{Heatmap of psuedodata.}
    \label{fig:enter-label}
\end{figure}

\subsection{Experimental Results}
The integration of GAN-generated data into our training process significantly improved the model's accuracy. Specifically, the testing loss was reduced from 62.93 cm to 49.295 cm, as shown in Fig 12. This reduction in loss indicates that the model's predictions are closer to the actual locations, demonstrating improved precision.

As illustrated in Fig 12, the predicted points (dots) are more accurately centered around their target points. This improvement is particularly notable in the left aisle, where the performance increased significantly. The enhanced accuracy in these regions can be attributed to the targeted data augmentation, which provided the model with more representative samples for training.

\begin{figure}
    \centering
    \includegraphics[width=1\linewidth, height=0.33\textheight]{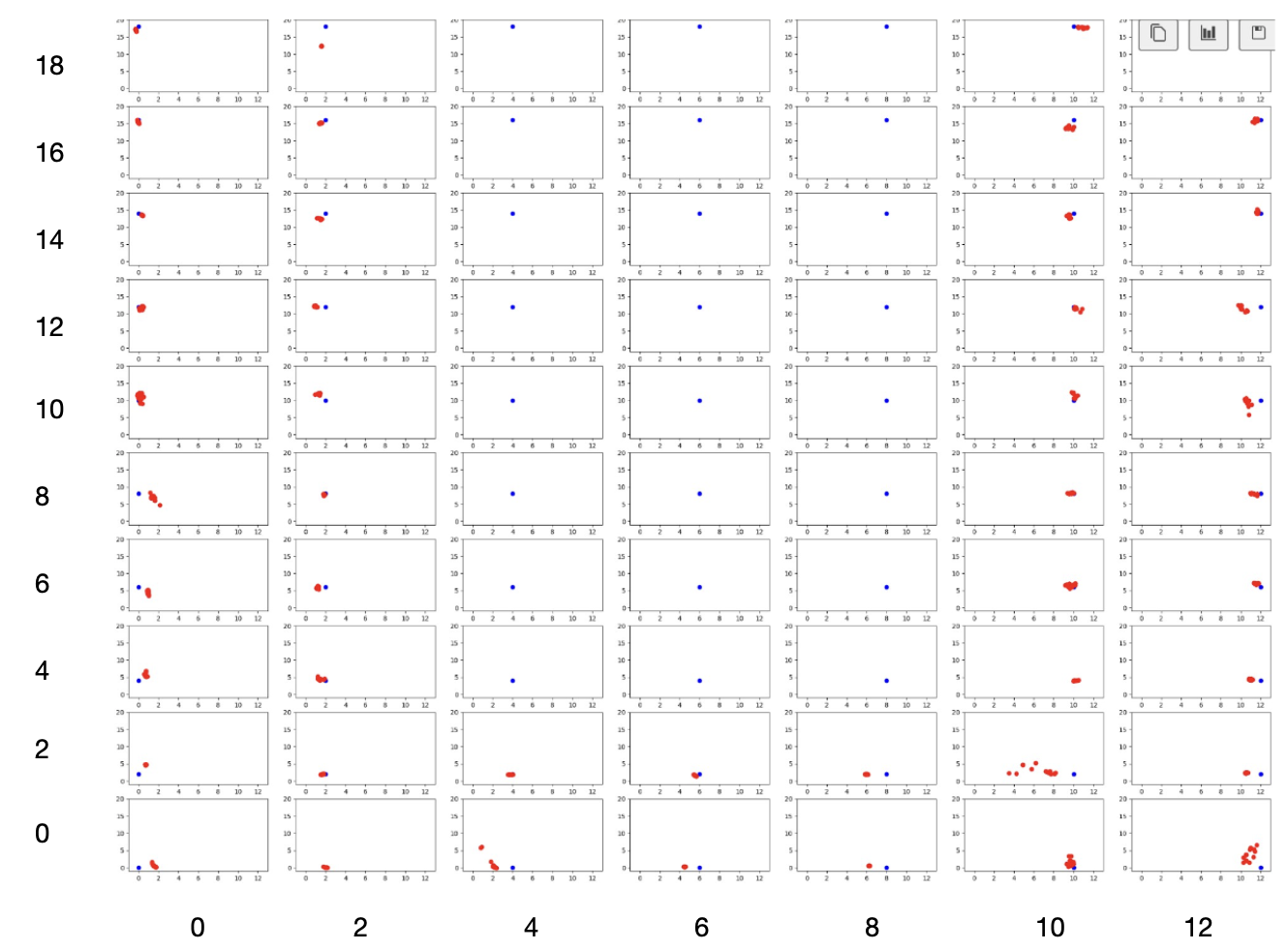}
    \caption{Performance of the model (with fake data).}
    \label{fig:enter-label}
\end{figure}

\section{\textbf{Challenges and Future Directions}}

During implementation we encountered several issues that limited prediction accuracy. Most stemmed from dataset characteristics and pre-processing choices.

\subsection{Intensity Normalization}

\textbf{Issue.}  Absolute light-intensity values vary across sessions, preventing the model from generalising.  
\textbf{Mitigation.}  Normalise each spectrum to a reference level before inference. For example, compute the intensity ratio between a fixed anchor point measured on the training day and on the prediction day, then scale all channels accordingly.  
\textbf{Next step.}  We plan to install coloured reference bulbs and recollect data in a controlled setting to obtain a stable normalisation baseline.

\subsection{Data Ordering}

\textbf{Issue.}  When new measurements were appended to the end of the CSV file, the ordering no longer followed the spatial \((x,y)\) grid, leading to inconsistent batches during training.  
\textbf{Mitigation.}  Maintain a deterministic ordering—e.g.\ sorted by coordinates or timestamp—whenever the dataset is updated. This guarantees that spatial neighbourhood information is preserved and learnt effectively.

\bigskip
By enforcing the condition, applying intensity normalisation, and standardising data order, we expect significant gains in robustness. Future work will evaluate these fixes on expanded datasets collected with controlled lighting.

\section{\textbf{Conclusions}}

By generating additional data with a GAN and integrating it into our training dataset, we addressed specific weaknesses in the model's predictions. This approach allowed us to fine-tune the model, leading to a substantial improvement in location prediction accuracy. The reduction in training loss and the improved visual alignment of predicted points with their targets validate the effectiveness of our strategy.

\section*{Acknowledgment}
The work was supported by Dr. F.-J.~Wu's projects funded by the National Science and Technology Council (NSTC) in Taiwan under Grant NSTC 113-2222-E-002 -001 -MY3, NSTC 112-2634-F-002-002-MBK, and NSTC 113-2221-E-002-202 -,  by the Ministry of Education (MOE) in Taiwan under the Yushan Fellow Program with the grant number NTU-112V1030-1 and NTU-113V1030-2, by NTU under the the project number NTU-NFG-113L7470, and by Deutsche Forschungsgemeinschaft (DFG, German Research Foundation) under the project number 511568981.

\addtolength{\textheight}{-12cm}


\bibliographystyle{IEEEtran}
\bibliography{sr}

\end{document}